\documentclass[twocolumn,pre,superscriptaddress,letterpaper]{revtex4}

\usepackage{dcolumn}
\usepackage{amsmath}

\usepackage{graphicx}
\usepackage{calrsfs}

\newcommand{\dd}{\mathrm{d}}

\newcommand{\av}[1]{\left\langle#1\right\rangle}

\newcommand{\mat}{\mathbf}
\renewcommand{\vec}{\mathbf}
\newcommand{\Ord}{\mathrm{O}}

\begin{document}

\title{Component structure and percolation in block models}
\author{Riccardo Franchi}
\affiliation{Department of Physics, University of Michigan, Ann Arbor, Michigan, USA}
\author{M. E. J. Newman}
\affiliation{Department of Physics, University of Michigan, Ann Arbor, Michigan, USA}
\affiliation{Center for the Study of Complex Systems, University of Michigan, Ann Arbor, Michigan, USA}

\begin{abstract}
The stochastic block model is a widely studied model of community structure in networks.  Here we study the component structure and percolation properties of networks generated from this model and its variants, using exact methods based on probability generating functions.  In particular, we derive expressions for the size of the giant component and the distribution of small components in such networks and for the size of the percolating cluster and position of the percolation threshold for both node and edge percolation, for the original stochastic block model and for its degree-corrected versions.  In passing, we also develop a mapping between generating functions for microcanonical and canonical block models that allows us to generalize results for the former to the latter with minimal effort.
\end{abstract}

\maketitle

\section{Introduction}
The stochastic block model is a widely used random graph model of community structure---a model of a network that has groups of densely connected nodes with sparser between-group connections, mimicking similar ``community structure'' observed in many real-world networks~\cite{HLL83,GN02,Fortunato10}.  The stochastic block model has been heavily researched as a tool for community detection, the inference of community structure in empirical network data, which is achieved by fitting the model to observed networks using maximum-likelihood or Bayesian methods~\cite{HLL83,WW87,SN97b,BC09,KN11a}.  In this paper we focus on two other aspects of the stochastic block model, its component structure and percolation properties.  The former tells us about connectivity in the network, typical component sizes, the size of the giant component if there is one, and how these things are affected by the presence and strength of community structure.  The latter tells us about the resilience of community structured networks as nodes or edges fail or are removed from the network.  Percolation phenomena are also closely connected to the behavior of epidemic models on networks and hence our results can shed light on the effects of community structure on the spread of disease over contact networks.

The stochastic block model comes in three main varieties, all of which we study here.  The first and original model, proposed by Holland et al.~\cite{HLL83}, is the simplest of the three.  It divides network nodes among a specified number of communities and then places edges independently at random between node pairs with probabilities that depend on the community membership of the nodes.  If edge probabilities are higher between nodes in the same community than nodes in different communities, this produces traditional ``assortative'' community structure with denser connections within groups than between them.  This model can be thought of as a community structured version of the classic random graph of Erd\H{o}s and R\'enyi~\cite{ER59,ER60}, and indeed the individual communities within the model take the form of Erd\H{o}s--R\'enyi networks.

Elegant though it is, however, the original stochastic block model has a significant flaw, in that it generates networks with a Poisson degree distribution within each community.  Most real-world networks have conspicuously non-Poisson degree distributions~\cite{AB02,ASBS00,BC19} and the inability of the stochastic block model to capture this fact severely degrades its performance, for example in community detection tasks.  This shortcoming is remedied by the degree-corrected stochastic block model~\cite{KN11a}, which introduces additional parameters controlling the expected degrees of nodes and thereby shaping the degree distribution.  In the same way that the original stochastic block model adds communities to the Erd\H{o}s--R\'enyi random graph, so the degree-corrected model adds them to the well known model of Chung and Lu~\cite{CL02a,CL02b} of a random graph with given expected degrees.

An alternative take on the degree-corrected block model is the micro\-canonical block model~\cite{Newman03c,Peixoto17}, which specifies exact degrees rather than expected ones.  In the same way that the normal degree-corrected model extends the Chung--Lu model, so the microcanonical model extends the standard configuration model of a network with given degree sequence.

There has, perhaps surprisingly, been little previous work on component structure and percolation in these models.  A model essentially equivalent to the microcanonical model was studied by one of us in~\cite{Newman03c}, where we gave a condition for the existence of a giant component in the network and a method for solving for the size of the giant component when there is one.  Bollob\'as et al.~\cite{BJR07} have studied a general class of ``inhomogeneous random graphs'' governed by an arbitrary kernel function.  Their methods could in principle be adapted to give results for the component structure of the stochastic block model and its variants, although this has not actually been done as far as we are aware.  Bujok et al.~\cite{BFF14} have studied component structure in the planted partition model, which is a special case of the standard stochastic block model.  Their work however focuses on finite-sized networks and the regime of a large number of small communities, where we focus on infinite networks and the more commonly studied regime of a small number of large communities.

The outline of this paper is as follows.  In Section~\ref{sec:components} we study the component structure of the various block models in order of increasing complexity, starting with the standard (non-degree-corrected) model, then progressing to the microcanonical and canonical degree-corrected models.  We derive formulas for the size of the giant component, if there is one, sizes of small components, and position of the phase transition at which a giant component appears.  In Section~\ref{sec:percolation} we look at node and edge percolation on the same three models, focusing particularly on the position of the percolation threshold, size of the percolating cluster, and sizes of the small clusters.  We also give a brief treatment of the case of nonuniform percolation, such as the removal of high-degree nodes from a network.  In Section~\ref{sec:conclusions} we give our conclusions.  Computer code for the numerical calculations presented is available at \url{github.com/riccardo-franchi/block-models} for download.

\section{Component structure}
\label{sec:components}
We begin by investigating the component structure of each of our three models, starting with the ordinary (not degree-corrected) stochastic block model, which presents a particularly simple case.

\subsection{Ordinary stochastic block model}
In the ordinary stochastic block model, $n$~network nodes are divided among $K$ groups such that there are $n_r$ nodes in group~$r$, with $r = 1 \ldots K$ and $\sum_r n_r = n$.  Edges are placed independently at random such that a pair of nodes, one in group~$r$ and the other in group~$s$, are connected by an edge with probability~$\omega_{rs}$.

The expected number of edges connecting a single node in group~$r$ to any of the nodes in group~$s$ is
\begin{equation}
    c_{rs} = n_s \omega_{rs},
    \label{eq:crs}
\end{equation}
which we can think of as a kind of ``group degree'' parameter---it is the expected number of group-$s$ neighbors the node has.  We assume that the network is sparse, such that both the values of the $c_{rs}$ and the fraction $n_r/n$ of the network occupied by group~$r$ tend to constants as the network becomes large, so that $n_r = \Ord(n)$ and $\omega_{rs} = \Ord(1/n)$.

If the edge probabilities $\omega_{rs}$ are sufficiently large, the network will contain a giant component---a connected component with size that scales with the size of the network, akin to the giant component in the standard random graph of Erd\H{o}s and R\'enyi~\cite{ER60,Bollobas01}.  Let $u_r$ be the probability that a randomly chosen node in group~$r$ is \emph{not} part of this giant component, if such a component exists.  (If it does not exist, then $u_r=1$.)  We can calculate $u_r$ by considering the probability that a node in group~$r$ is not connected to the giant component via a specific other node in group~$s$.  There are two ways this can happen.  First, there might be no edge between the two nodes, which occurs with probability $1-\omega_{rs}$, or second, there might be an edge (probability~$\omega_{rs}$) but the node in group~$s$ is itself not in the giant component~(probability~$u_s$).  Putting these together, the total probability is $1-\omega_{rs} + \omega_{rs}u_s$ and the probability that a node in group~$r$ is not connected to the giant component via \emph{any} node in group~$s$ is this quantity to the power of~$n_s$.  Taking a product over all groups, we then have
\begin{equation}
    u_r = \prod_s (1 - \omega_{rs} + \omega_{rs} u_s)^{n_s}
        = \prod_s \biggl[ 1 - {c_{rs}\over n_s} (1-u_s) \biggr]^{n_s},
\end{equation}
where we have used Eq.~\eqref{eq:crs}.  Taking logs of both sides, we have
\begin{align}
    \ln u_r &= \sum_s n_s \log \biggl[ 1 - {c_{rs}\over n_s} (1-u_s) \biggr] \nonumber\\
    &= - \sum_s \bigl[ c_{rs} (1 - u_s) + \Ord(1/n_s) \bigr].
\end{align}
The sub-leading terms can be neglected in the limit of large network size where $n_s\to\infty$ and, taking the exponential again, we have
\begin{equation}
    u_r = e^{-\sum_s c_{rs}(1-u_s)}.
\end{equation}
Defining $S_r = 1 - u_r$ to be the fraction of group-$r$ nodes in the giant component, we then have
\begin{equation}
  S_r = 1 - e^{-\sum_s c_{rs} S_s}.
\label{eq:gc_SBM}
\end{equation}
Then the overall size of the entire giant component is given by the average
\begin{equation}
    S = {1\over n} \sum_r n_r S_r.
\end{equation}

Equation~\eqref{eq:gc_SBM} is a natural generalization of the corresponding result $S = 1 - e^{-cS}$ for the Erd\H{o}s--R\'enyi random graph and reduces to that result in the case of a single group.  In practice one cannot solve for the $S_r$ in closed form, but it is straightforward to solve for them numerically, by simple iteration.  One guesses initial values for the~$S_r$ (for instance at random) and then iterates Eq.~\eqref{eq:gc_SBM} to convergence.

If there is no giant component in the network, then $S_r$ will converge to zero for all~$r$.  Otherwise it will converge to some nonzero value.  Thus a condition for the existence of a giant component is that the point $S_r=0$ should not be an attracting (stable) fixed point of the iteration.  Stability can be determined by linearizing.  Writing the iteration as $S_r' = 1 - e^{-\sum_s c_{rs} S_s}$ and expanding for small~$S_r$, we have $S_r' = \sum_s c_{rs} S_s + \Ord(S^2)$, or in vector notation $\vec{s}' = \mat{C}\vec{s}$ to leading order, where $\mat{C}$ is the matrix with elements~$c_{rs}$.  The vector $\vec{s}$ will grow and there will be a giant component in the network if and only if the leading eigenvalue of~$\mat{C}$ is greater than unity.

As an example, consider the so-called planted partition model, which is the special case where all groups have the same size $n_r  = n/K$ and there are only two edge probabilities, one for in-group connections and one for between-group connections:
\begin{equation}
    \omega_{rs} = \biggl\lbrace\begin{array}{ll}
        \omega_\text{in} & \quad\text{if $r=s$,} \\
        \omega_\text{out} & \quad\text{if $r\ne s$.}
    \end{array}
\end{equation}
Then, by symmetry, all groups are equivalent and hence have the same value of~$S_r$.  Denoting this value by~$S$, we have $\sum_s c_{rs} S_r = S \sum_s c_{rs} = cS$, where $c$ is the overall average degree of a node, and Eq.~\eqref{eq:gc_SBM} reduces to $S = 1 - e^{-cS}$, which is the same as the standard result for the Erd\H{o}s--R\'enyi random graph.  Hence for this special case the size of the giant component and the condition for its existence are the same as in that model.

One can calculate many other properties for the ordinary stochastic block model, but we will not pursue them here since the model is considered somewhat unrealistic and has been largely superseded in recent work by its degree-corrected variant, which we study next.

\subsection{Microcanonical degree-corrected block model}
\label{sec:microcanonical}
Turning to the degree-corrected block model, we first look at the microcanonical version~\cite{Peixoto17}.  In this model $n$ nodes are again assigned to $K$ groups of sizes~$n_r$, but now in addition we specify the exact degree~$k_i$ of each node~$i$, along with the number (not the probability) of edges between each pair of groups.  Specifically, we define a $K\times K$ mixing matrix~$\mat{M}$ with integer elements $m_{rs}$ equal to the number of ends of edges in group~$r$ whose other end is in group~$s$.  Equivalently, $m_{rs}$ is equal to the number of edges between groups $r$ and~$s$, or twice that number when $r=s$.

Given these numbers, we now create the network by giving each node~$i = 1\ldots n$ a total of $k_i$ edge stubs and then matching stubs randomly in pairs to create edges while satisfying the constraints imposed by the mixing matrix---technically, the network is a uniform draw from the set of all matchings that respect the specified values of~$m_{rs}$.

It will be useful to consider the degree distribution of the network, and to regard it as a composite of the distributions in each of the $K$ groups.  Let $p^{(r)}_k$ be the probability that a randomly chosen node in group~$r$ has degree~$k$:
\begin{equation}
    p^{(r)}_k = {1\over n_r} \sum_{i\in r} \delta_{k_ik},
\end{equation}
where the sum is over all nodes~$i$ belonging to group~$r$ and $\delta_{ij}$ is the Kronecker delta.  A related and important quantity is the so-called excess degree distribution:
\begin{equation}
    q^{(r)}_k = {(k+1) p^{(r)}_{k+1}\over\sum_k k p^{(r)}_k},
    \label{eq:excess_dcsbm}
\end{equation}
which is the probability that, upon following an edge to reach a node in group~$r$, we find that node to have exactly~$k$ other edges attached to it---its excess degree is~$k$.

As in the previous section, we will assume the network to be sparse, which in this context we take to mean that the degree distributions~$p^{(r)}_k$ are constant in the limit of large network size, so that all moments are defined, average degree is constant, and the average probability of an edge is of order~$1/n$.

\subsubsection{Giant component}
\label{sec:gc_DCSBM}
Armed with these definitions, we can now calculate a variety of properties of networks generated by the microcanonical model, starting with the size of the giant component.  Consider an edge attached to a node in group~$r$ and let~$u_r$ be the probability that such an edge does not lead to the giant component.  Because the model is defined as a random matching of edge stubs, this probability does not depend on which node we start from in group~$r$---every edge in~$r$ has the same probability of attaching to any given stub and hence the same probability of leading to the giant component.

Suppose that this particular edge connects to a node in group~$s$ that has excess degree~$k$.  The probability that the latter node is not itself in the giant component is equal to the probability that none of its other edges connect to the giant component, which is~$u_s^k$.  The average of this probability over all possible values of the excess degree is $\sum_k q^{(s)}_k u_s^k$, which takes the form of a probability generating function, which we denote by
\begin{equation}
    g^{(s)}_1(z) = \sum_k q^{(s)}_k z^k.
    \label{eq:g1_dcsbm}
\end{equation}
At the same time, the probability that the original edge connected to a node in group~$s$ is equal to
\begin{equation}
    {m_{rs}\over\sum_s m_{rs}} = {m_{rs}\over\kappa_r},
\end{equation}
where
\begin{equation}
    \kappa_r = \sum_s m_{rs} = \sum_{i\in r} k_i,
    \label{eq:kappar}
\end{equation}
is the total number of edge ends connected to nodes in group~$r$, or equivalently the sum of the degrees of all group-$r$ nodes.  Summing over all groups, we now have
\begin{equation}
  u_r = {1\over\kappa_r} \sum_s m_{rs} \sum_k q^{(s)}_k\, u_s^k = {1\over\kappa_r} \sum_s m_{rs}\, g^{(s)}_1(u_s).
\label{eq:u_dcsbm}
\end{equation}
(Note that Eq.~\eqref{eq:kappar} also imposes a constraint on~$m_{rs}$, fixing its row and column sums once the node degrees are specified.)

If we can solve the system of equations in~\eqref{eq:u_dcsbm} for the quantities~$u_r$, then the probability that a node of degree~$k$ in group~$r$ is not in the giant component is simply~$u_r^k$.  Averaging over the degree distribution within the group, the average probability that a group-$r$ node is not in the giant component is $\sum_k p^{(r)}_k u_r^k$, which again takes the form of a probability generating function, which we denote by
\begin{equation}
    g^{(r)}_0(z) = \sum_k p^{(r)}_k z^k.
    \label{eq:g0r}
\end{equation}
Note that the two generating functions~$g^{(r)}_0$ and $g^{(r)}_1$ are not really independent, since the degree distribution and excess degree distribution are related by Eq.~\eqref{eq:excess_dcsbm}.  Specifically, we can show that
\begin{equation}
    g^{(r)}_1(z) = {\dd g^{(r)}_0/\dd z\over(\dd g^{(r)}_0/\dd z)_{z=1}}.
\end{equation}

Now the probability~$S_r$ that a group-$r$ node is in the giant component is
\begin{equation}
    S_r = 1 - g^{(r)}_0(u_r),
    \label{eq:Sr_dcsbm}
\end{equation}
and the average probability for any node in network, which is also the expected size of the giant component as a fraction of the system size, is
\begin{equation}
    S = {1\over n} \sum_r n_r S_r = 1 - {1\over n} \sum_r n_r g^{(r)}_0(u_r).
\end{equation}

The equations~\eqref{eq:u_dcsbm} can be solved by simple iteration starting from any suitable initial values, such as random values.  If there is no giant component in the network then necessarily this iteration will converge to $u_r=1$ for all~$r$---recall that $u_r$ is the probability that an edge does \emph{not} lead to the giant component, so $u_r=1$ implies that no edge leads to the giant component and hence there is no giant component.  If there is a giant component then the iteration will converge to some other solution.  Thus there is a giant component if the point $u_r=1$ is unstable, which we can determine by linearizing.  Setting $u_r = 1 - \epsilon_r$ in Eq.~\eqref{eq:u_dcsbm} and expanding in the small quantity~$\epsilon_r$, the iteration takes the form
\begin{align}
    1 - \epsilon_r' &= {1\over\kappa_r} \sum_s m_{rs} g^{(s)}_1(1-\epsilon_s) \nonumber\\
    &= {1\over\kappa_r} \sum_s m_{rs} \biggl[ 1 - \epsilon_s \biggl( {\dd g^{(s)}_1\over\dd z} \biggr)_{z=1} + \Ord(\epsilon_s^2) \biggr],
    \label{eq:epsilon_dcsbm}
\end{align}
which can be written to leading order as
\begin{equation}
    \epsilon_r' = \sum_s J_{rs} \epsilon_s,
    \label{eq:stability_dcsbm}
\end{equation}
where $J_{rs}$ is an element of the $K\times K$ Jacobian matrix~$\mat{J}$:
\begin{equation}
    J_{rs} = {m_{rs}\over\kappa_r} \tilde{c}_s
\end{equation}
and we have defined the quantity
\begin{equation}
    \tilde{c}_r = \biggl( {\dd g^{(r)}_1\over\dd z} \biggr)_{z=1} \!\! = \>\sum_k k q^{(r)}_k,
    \label{eq:tildec}
\end{equation}
which is the average excess degree of a node in group~$r$.

The value of $\epsilon_r$ will grow under iteration of~\eqref{eq:stability_dcsbm}, and hence the solution at $u_r=1$ is unstable, if and only if the leading eigenvalue of $\mat{J}$ is greater than one.  This defines the condition for the existence of a giant component in the network, which is a generalization of the condition, given by Molloy and Reed~\cite{MR95}, for the existence of a giant component in the standard configuration model.

As an example, consider a network with two communities of $\frac12 n$ nodes each, having geometric (i.e.,~exponential) degree distributions
\begin{equation}
    p^{(1)}_k = (1-a_1)\,a_1^k,\qquad
    p^{(2)}_k = (1-a_2)\,a_2^k,
    \label{eq:micro_geometric}
\end{equation}
with constant parameters $a_1,a_2<1$.  Via Eq.~\eqref{eq:excess_dcsbm} this implies excess degree distributions
\begin{equation}
    q^{(1)}_k = (1-a_1)^2 (k+1)\,a_1^k,\quad
    q^{(2)}_k = (1-a_2)^2 (k+1)\,a_2^k,
\end{equation}
and hence
\begin{equation}
    \tilde{c}_r = {2a_r\over1-a_r},\qquad \kappa_r = \tfrac12 n {a_r\over1-a_r}.
\end{equation}
In order to satisfy Eq.~\eqref{eq:kappar} we must have a mixing matrix $\mat{M} = [m_{rs}]$ of the form
\begin{equation}
    \mat{M} = \begin{pmatrix}
        \kappa_1 - t & t \\
        t & \kappa_2 - t \\
    \end{pmatrix}
\end{equation}
for some integer~$t\le\kappa_1,\kappa_2$.  Then it is straightforward to show that the Jacobian is
\begin{equation}
\setlength{\arraycolsep}{6pt}
    \mat{J} = 2\begin{pmatrix}
        c_1 - x & xc_2/c_1 \\
        xc_1/c_2 & c_2 - x \\
    \end{pmatrix},
    \label{eq:micro_jacobian}
\end{equation}
where $c_r = a_r/(1-a_r)$ is the average degree of group~$r$ and $x = 2t/n$ is a non-negative real number which must be less than the smaller of $c_1$ and~$c_2$.  The leading eigenvalue is then given by standard formulas to be
\begin{equation}
    \lambda = c_1+c_2-2x + \sqrt{(c_1-c_2)^2+4x^2}.
\end{equation}
A giant component exists if and only if this value is greater than one.  For instance, if $x=0$, so that there are no between-group edges, then we find $\lambda = c_1 + c_2 + |c_1-c_2| = \max(2c_1,2c_2)$ and hence there will be a giant component if either $c_1$ or $c_2$ is greater than~$\frac12$, which is equivalent to the condition that there is a giant component in either one of the two groups (or both) under the standard Molloy-Reed criterion~\cite{MR95}.

On the other hand, if $x=\frac14$, thereby introducing some between-group edges, then we have
\begin{equation}
    \lambda = c_1 + c_2 - \tfrac12
    + \sqrt{(c_1-c_2)^2+\tfrac14}.
    \label{eq:geom_gc}
\end{equation}
Setting $\lambda>1$ we find (after some manipulation) that there is a giant component if $c_1+c_2>1$ or $(4c_1-3)(4c_2-3) < 1$, but excluding points with $c_1,c_2\le\frac14$ (because $c_1$ and $c_2$ must be greater than~$x=\frac14$, otherwise the number of in-group edges would be negative---see Eq.~\eqref{eq:micro_jacobian}).

\begin{figure}
    \begin{center}
        \includegraphics[width=\columnwidth]{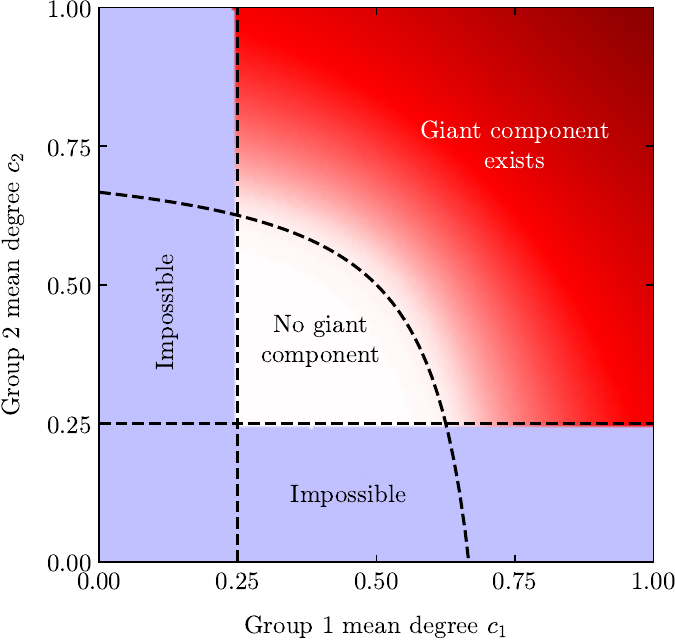}
    \end{center}
    \caption{Phase diagram for the network with two equally sized groups and geometric degree distributions with means $c_1$ and~$c_2$, for the case $x=\frac14$ of Eq.~\eqref{eq:geom_gc}.  A giant component exists only in the marked region at top right.  Below this, in the region labeled ``No giant component,'' the network has small components only, and in the regions labeled ``Impossible'' there can be no network at all, because it would require a negative number of in-group edges.  The shading shows the results of simulations, averaged over 100 networks of 100\,000 nodes each.  Shades of red indicate the size of the largest component; blue indicates parameter regimes where we are unable to create a network because of negative edge counts.}
    \label{fig:phasediag}
\end{figure}

Figure~\ref{fig:phasediag} shows a phase diagram of the system, with dashed lines indicating the boundaries of the various behaviors.  A giant component exists in the top-right of the plot above the curved line.  Below the line is a region where there are small components only but no giant component.  For parameter values outside of the two straight dashed lines at $c_1,c_2=\frac14$ there are no valid networks.  The shading in the figure shows the results of direct simulations of microcanonical block models with the same parameter values, which are in good agreement with the theory.  Shades of red indicate the size of the largest component and we can see that indeed there is a large component above the curve, but not below (although it becomes smaller as we approach the line).  Blue indicates areas where it was not possible to generate a network because the number of in-group edges would have to be negative.

\subsubsection{Small components}
\label{sec:small}

We can calculate a range of further properties of the microcanonical model, such as for example the size distribution of the small components.  As in the conventional configuration model, the small components take the form of trees with high probability in the limit of large network size~$n$ because the network is locally tree-like, and this property allows us to calculate their size.

Let $\pi^{(r)}_t$ be the probability that a node in group~$r$ belongs to a small component of size~$t$ and let us define a generating function for this probability thus:
\begin{equation}
    h^{(r)}_0(z) = \sum_t \pi^{(r)}_t z^t.
\end{equation}
Similarly, let $\rho^{(r)}_t$ be the probability that an edge from a node in group~$r$ connects to a small (sub)component of $t$ nodes and let us define the corresponding generating function
\begin{equation}
    h^{(r)}_1(z) = \sum_t \rho^{(r)}_t z^t.
\end{equation}
Then, because of the tree-like form of the small components, the number of nodes reachable from a certain node in group~$r$ is the sum of the numbers reachable along each of its edges individually and, following arguments of~\cite{NSW01,Newman18c} for the configuration model, if the node has degree~$k$ then the probability distribution of the reachable number has generating function $[h^{(r)}_1(z)]^k$.  Averaging over~$k$, the average generating function is
\begin{equation}
    \sum_k p^{(r)}_k [h^{(r)}_1(z)]^k
      = g^{(r)}_0\bigl( h^{(r)}_1(z) \bigr),
      \label{eq:h0_step}
\end{equation}
where we have used the definition of $g^{(r)}_0$ from Eq.~\eqref{eq:g0r}.  The size of the component a node belongs to is equal to the number of reachable nodes plus one for the node itself, which adds one more factor of~$z$ to the generating function and hence we have
\begin{equation}
    h^{(r)}_0(z) = z g^{(r)}_0\bigl( h^{(r)}_1(z) \bigr).
    \label{eq:h0_dcsbm}
\end{equation}
Then the overall generating function for the size of the component to which any node belongs is given by the average
\begin{equation}
    h_0(z) = {1\over n} \sum_r n_r h^{(r)}_0(z).
    \label{eq:h0_overall}
\end{equation}

The coefficients of this generating function give us the overall distribution of component sizes, but we still need to calculate~$h^{(r)}_1(z)$, which we do by an argument similar to the one for Eq.~\eqref{eq:h0_step}.  If an edge leads to a node in group~$s$ with excess degree~$k$ then the probability distribution of the number of further nodes reachable from that node has generating function $[h^{(s)}_1(z)]^k$.  Averaging over $k$ then gives an average generating function
\begin{equation}
    \sum_k q^{(s)}_k [h^{(s)}_1(z)]^k
      = g^{(s)}_1\bigl( h^{(s)}_1(z) \bigr).
\end{equation}
Counting the node itself adds an extra factor of~$z$ and then we average over all groups~$s$ to get the full generating function for the number of nodes reachable along an edge from group~$r$:
\begin{equation}
    h^{(r)}_1(z) = {z\over\kappa_r} \sum_s m_{rs} g^{(s)}_1\bigl( h^{(s)}_1(z) \bigr).
    \label{eq:h1_dcsbm}
\end{equation}

Armed with Eqs.~\eqref{eq:h0_dcsbm}, \eqref{eq:h0_overall}, and~\eqref{eq:h1_dcsbm} we can now perform various calculations.  The average size of the small component to which a node in group~$r$ belongs, for instance, is given by the derivative of the generating function~$h^{(r)}_0$:
\begin{equation}
    \av{t}_r = {1\over h^{(r)}_0(1)} \biggl( {\dd h^{(r)}_0\over\dd z} \biggr)_{z=1}.
\end{equation}
Note that the normalizing factor of $1/h^{(r)}_0(1)$ is needed because $\pi^{(r)}_t$ is not a conventionally normalized probability distribution---the total probability of belonging to a small component of any size is not equal to~1, but to the fraction of group~$r$ not occupied by the giant component:
\begin{equation}
    \sum_t \pi^{(r)}_t = h^{(r)}_0(1) = 1 - S_r.  
\end{equation}
Similarly $h^{(r)}_1(1)$ is equal to the total probability of reaching a small component when following an edge from group~$r$, which is also the quantity we called $u_r$ in our previous analysis:
\begin{equation}
    h^{(r)}_1(1) = u_r.
\end{equation}
Applying Eq.~\eqref{eq:h0_dcsbm}, we now have
\begin{equation}
    \av{t}_r = 1 + {1\over g^{(r)}_0(u_r)} \biggl( {\dd g^{(r)}_0\over \dd z} \biggr)_{z=u_r} \biggl( {\dd h^{(r)}_1\over\dd z} \biggr)_{z=1}.
    \label{eq:avt1}
\end{equation}
Similarly, differentiating Eq.~\eqref{eq:h1_dcsbm} gives
\begin{equation}
    \biggl( {\dd h^{(r)}_1\over\dd z} \biggr)_{z=1} \!\!
      = u_r + {1\over\kappa_r} \sum_s m_{rs} \biggl( {\dd g^{(s)}_1\over\dd z} \biggr)_{z=u_s} \biggl( {\dd h^{(s)}_1\over\dd z} \biggr)_{z=1},
\end{equation}
which can be solved numerically by iteration, or as a system of linear equations, and then substituted into~\eqref{eq:avt1} to give a value for~$\av{t}_r$.

We can go further than calculating just the average component size, however.  We can also compute the entire distribution of small components from the generating functions~$h^{(r)}_0(z)$.  For the corresponding calculation on the configuration model it is possible to solve for the distribution in closed form by an application of the Lagrange inversion theorem~\cite{Newman07}, but no equivalent result exists in the present case, as far as we are able to determine.  One can proceed numerically, however, and we describe two ways to perform the calculation.

The first method involves iterating Eq.~\eqref{eq:h1_dcsbm} to calculate~$h^{(r)}_1(z)$ to finite order in~$z$.  By keeping careful track of the coefficients of the generating function this calculation can be done in a numerically exact manner in a finite number of iterations.  The crucial observation is that Eq.~\eqref{eq:h1_dcsbm} has a leading factor of $z$ on the right-hand side, which physically represents the fact that one always reaches at least one node when one follows an edge.  This means that $h^{(r)}_1(z) = \Ord(z)$ and, expanding the generating function~$g^{(r)}_1$ in Eq.~\eqref{eq:h1_dcsbm} according to its definition~\eqref{eq:g1_dcsbm}, we have
\begin{align}
    h^{(r)}_1(z) &= z g^{(r)}_1\bigl( h^{(r)}_1(z) \bigr) \nonumber\\
    &= z \bigl[ q^{(r)}_0 + q^{(r)}_1 h^{(r)}_1(z) + q^{(r)}_2 [ h^{(r)}_1(z) ]^2 + \ldots \bigr] \nonumber\\
    &= q^{(r)}_0 z + q^{(r)}_1 \Ord(z^2) + q^{(r)}_2 \Ord(z^3) + \ldots
    \label{eq:h1_series}
\end{align}
From this expression we can see that the coefficient of $z^t$ in the generating function~$h^{(r)}_1(z)$ on the left contains contributions only from the first~$t$ terms in the expansion, and hence only from terms up to $z^{t-1}$ in the generating function~$h^{(r)}_1(z)$ on the right.  This means that if $h^{(r)}_1(z)$ on the right is exact up to terms in $z^{t-1}$ then $h^{(r)}_1(z)$ on the left is exact up to terms in~$z^t$.

We use these observations to construct a numerical scheme in which, rather than recording the actual value of~$h^{(r)}_1(z)$ for any $z$ we record the coefficients~$\rho^{(r)}_t$ in its Taylor series expansion.  Initially we set the first two coefficients to $\rho^{(r)}_0 = 0$ and $\rho^{(r)}_1 = q^{(r)}_0$ which, following Eq.~\eqref{eq:h1_series} are exactly correct.  Then we apply Eq.~\eqref{eq:h1_dcsbm} once to calculate a new value, which will be exact up to the coefficient~$\rho^{(r)}_2$.  Then we iterate the process, applying Eq.~\eqref{eq:h1_dcsbm} repeatedly, which gives one more coefficient on each iteration.  To compute the first $t$ coefficients exactly one need only keep track of those coefficients and only perform $t-1$ iterations.  The calculation can be performed conveniently using symbolic manipulation software which can keep track of the series expansion directly, or more quickly but also more laboriously using custom code that explicitly stores the coefficients and computes sums and products of polynomials using Horner's method~\cite{CLRS01}.  Once we have the values of the $\rho^{(r)}_t$ then a single application of Eq.~\eqref{eq:h0_dcsbm} gets us the values of the probabilities~$\pi^{(r)}_t$.

\begin{figure}[t]
  \begin{center}
  \includegraphics[width=\columnwidth]{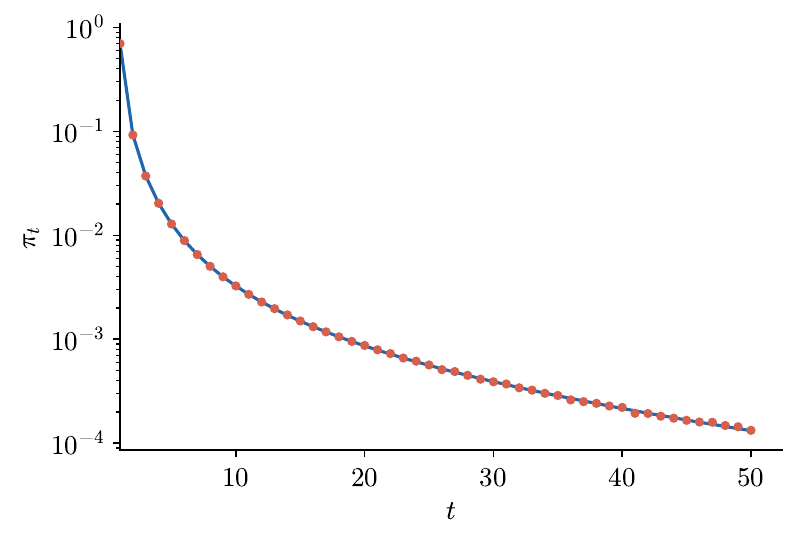}
  \end{center}
  \caption{Distribution of component sizes for a microcanonical block model consisting of two equally-sized groups with geometric degree distributions $p^{(r)}_k = (1-a_r)\,a_r^k$, parameters $a_1=0.2$ and $a_2=0.4$, and mixing parameter ${m_{12}/\kappa_1} = 0.8$.  The solid line represents the exact solution, while the points are measurements of $\pi_t$ averaged over 100 computer-generated networks of 4~million nodes each.}
  \label{fig:small_components}
\end{figure}

Figure~\ref{fig:small_components} shows an example application of this method to a network with two groups, each with a geometric degree distribution as in Eq.~\eqref{eq:micro_geometric}, which gives generating functions
\begin{equation}
    g^{(r)}_0(z) = {1-a_r\over1-a_rz}, \qquad
    g^{(r)}_1(z) = \biggl( {1-a_r\over1-a_rz} \biggr)^2.
    \label{eq:geometric_g}
\end{equation}
The solid curve in the figure shows the numerically exact result for the distribution of small component sizes, while the discrete points show the results of direct simulations of the microcanonical model in which component sizes are measured using traditional breadth-first search and the results averaged over 100 realizations of the network.  As the figure shows, agreement is good between the exact computation and simulations.

A simpler though somewhat less accurate way to solve for the component sizes is to use the Cauchy derivative formula to calculate the series coefficients of the generating function~$h^{(r)}_0$:
\begin{equation}
  \pi^{(r)}_t = \frac{1}{t!} \biggl( {\dd^t h^{(r)}_0\over\dd z^t} \biggr)_{z=0}
           = \frac{1}{2\pi i} \oint \frac{h^{(r)}_0(z)}{z^{t+1}}\,\dd z,
\end{equation}
where the integral is performed counterclockwise around any contour that encircles the origin once but encloses no poles in~$h^{(r)}_0(z)$.  Here we choose the contour to lie on the unit circle $z = e^{i\theta}$, which means that
\begin{equation}
  \pi^{(r)}_t = \frac{1}{2\pi} \int_0^{2\pi} h^{(r)}_0\bigl(e^{i\theta}\bigr)\, e^{-i\theta t}\,\dd\theta.
\end{equation}
We approximate the integral numerically using the trapezoidal rule on $N$ points $\theta_n = 2\pi n/N$, giving
\begin{equation}
  \pi^{(r)}_t \simeq \frac{1}{N} \sum_{n=0}^{N-1} h^{(r)}_0\bigl(e^{i\theta_n}\bigr)\, e^{-2\pi itn/N},
\end{equation}
which takes the form of a discrete Fourier transform of the complex function $h^{(r)}_0\bigl(e^{i\theta_n}\bigr)$, which can be computed rapidly using a fast Fourier transform.  Applying this approach to the same example system as in Fig.~\ref{fig:small_components} we find good agreement with our previous results.

\subsection{Canonical model}
\label{sec:canonical}
The canonical degree-corrected block model (as opposed to the microcanonical one) is the original and most widely studied version of the model, in which one fixes not the number but the probability of edges within and between groups~\cite{KN11a}.  It turns out that the component structure of this model is relatively easy to derive given the results for the microcanonical model in the previous section, by using a specific mapping of the generating functions.

We illustrate this process first for the configuration model, for which an analogous argument can be made but the developments are simpler.  In the language of this paper, the configuration model is a microcanonical model, a random graph in which the exact degree of each node is specified.  The canonical equivalent is the model of Chung and Lu~\cite{CL02a,CL02b}, which specifies a single scalar parameter~$c_i$ for each node and then connects each pair of nodes $i,j$ by an edge with probability $p_{ij} = c_i c_j/2m$, where $m$ is the expected number of edges in the whole network, given by $2m = \sum_i c_i$.  The expected degree of node~$i$ within this model is
\begin{equation}
\sum_j p_{ij} = {c_i\over2m} \sum_j c_j = c_i,
\end{equation}
but the actual degree is not fixed by the generative process.  Instead the degree follows a Poisson distribution with mean~$c_i$.

An equivalent way to generate a network from the Chung--Lu model is first to draw a degree for each node from the appropriate Poisson distribution, then build a microcanonical configuration model network with those degrees.  This process generates each possible network with the same probability as the original Chung--Lu model, which means that one can calculate average properties of the Chung--Lu model by solving for the properties of the configuration model and then averaging over all possible degree sequences, each weighted with the appropriate Poisson probability.

The large-scale properties of the configuration model, however, such as the component structure that is our focus here, depend in the limit of large graph size only on the overall distribution of degrees~$p_k$ and not on the specific degree sequence~\cite{MR98,NSW01}, so it is not actually necessary to separately solve the model for each degree sequence and perform the average.  An equivalent and simpler procedure is to find the overall distribution of degrees implied by the Poisson distributions on the individual nodes and then solve the configuration once only for that implied distribution.

The latter distribution is straightforward to calculate.  Suppose that, in the limit of large network size, the probability density, over the whole network, of the node parameters~$c_i$ is~$p(c)$.  Given a particular~$c$, the probability of having degree~$k$ is given by the Poisson distribution
\begin{equation}
P(k|c) = {c^k\over k!} e^{-c},
\end{equation}
and hence the overall probability~$p_k$ of drawing a degree~$k$---which is also the final degree distribution of the whole network---is
\begin{equation}
    p_k = \int_0^\infty P(k|c)\,p(c) \>\dd c = \int_0^\infty {c^k\over k!} e^{-c}\,p(c) \>\dd c.
\end{equation}

Now we solve for the component structure of the configuration model with this degree distribution using standard generating-function methods~\cite{NSW01,Newman18c}.  The probability generating function for the degree distribution is
\begin{align}
    g_0(z) &= \sum_{k=0}^\infty p_k z^k = \int_0^\infty p(c) \sum_{k=0}^\infty {(cz)^k\over k!} e^{-c} \>\dd c
    \nonumber\\
    &= \int_0^\infty p(c)\,e^{-c(1-z)} \>\dd c = \mathcal{L}\{p\}(1-z),
    \label{eq:replace0}
\end{align}
where $\mathcal{L}\{p\}(z)$ is the Laplace transform of the probability distribution~$p(c)$.

The average degree of the network is
\begin{align}
    \av{k} &= \sum_{k=0}^\infty k p_k = \int_0^\infty c p(c) \sum_{k=1}^\infty {c^{k-1}\over (k-1)!} e^{-c} \>\dd c \nonumber\\
    &= \int_0^\infty c p(c) \>\dd c = \av{c},
\end{align}
and hence the excess degree distribution~\cite{NSW01,Newman18c} is
\begin{equation}
    q_k = {(k+1) p_{k+1}\over\av{k}} = {1\over\av{c}} \int_0^\infty {c^{k+1}\over k!} e^{-c}\,p(c) \>\dd c,
\end{equation}
and the corresponding generating function is
\begin{align}
    g_1(z) &= \sum_{k=0}^\infty q_k z^k
    = {1\over\av{c}} \int_0^\infty c p(c)\sum_{k=0}^\infty {(cz)^k\over k!} e^{-c} \>\dd c \nonumber\\
    &= {1\over\av{c}} \int_0^\infty c p(c)\,e^{-c(1-z)} \>\dd c = \mathcal{L}\{q\}(1-z),
    \label{eq:replace1}
\end{align}
where $\mathcal{L}\{q\}(z)$ is the Laplace transform of
\begin{equation}
q(c) = {c p(c)\over\av{c}},
\end{equation}
which plays the role of the excess distribution for the degree parameter~$c$ in the Chung--Lu model.

Thus the two fundamental generating functions $g_0$ and $g_1$ for the degree distribution and excess degree distribution are replaced by Laplace transforms of the corresponding distributions in the Chung--Lu model.  The Laplace transform is the natural generalization of a generating function to a continuous probability density and arises in similar situations elsewhere in the networks literature~\cite{CKN21}.  Once we make this replacement, the rest of the calculation follows normal lines for the (microcanonical) configuration model, with things like the criterion for existence of a giant component and the size of the giant and small components given by standard formulas.  For instance, the size~$S$ of the giant component, which is given in the configuration model by the solution of the equations
\begin{equation}
    S = 1 - g_0(u), \qquad u = g_1(u),
\end{equation}
is given in the Chung--Lu model by
\begin{equation}
    S = 1 - \mathcal{L}\{p\}(1-u), \qquad u = \mathcal{L}\{q\}(1-u).
\end{equation}

The same reasoning can now be applied to the canonical version of the degree-corrected stochastic block model.  In this model the probability~$p_{ij}$ of an edge between nodes $i$ and $j$ belonging to groups~$g_i$ and $g_j$ is
\begin{equation}
    p_{ij} = {c_i c_j\over2m}\,\omega_{g_ig_j},
\end{equation}
which modifies the Chung--Lu model with the group-dependent factor~$\omega_{g_ig_j}$ to create modular structure.  The quantity~$m$ is again given by $2m = \sum_i c_i$ and we can arrange that $c_i$ represents the expected degree of node~$i$ by ensuring that $c_i = \sum_j p_{ij} = (c_i/2m) \sum_j c_j \omega_{g_ig_j}$, or
\begin{equation}
    \sum_j c_j \omega_{rg_j} = 2m
    \label{eq:condition1}
\end{equation}
for all~$r$.  Or we can write
\begin{align}
    \sum_j c_j \omega_{rg_j} &= \sum_j c_j \sum_s \delta_{sg_j} \omega_{rs}
    = \sum_s \omega_{rs} \sum_j c_j \delta_{sg_j} \nonumber\\
    &= \sum_s \omega_{rs} \kappa_s,
\end{align}
where $\kappa_r$ is now the sum of $c_i$ for all nodes in group~$r$---the canonical equivalent of the quantity of the same name for the microcanonical model.  Thus the condition of Eq.~\eqref{eq:condition1} can also be written
\begin{equation}
    \sum_s \omega_{rs} \kappa_s = 2m.
\end{equation}
This equation plays a similar role to Eq.~\eqref{eq:kappar} for the microcanonical model.  We will assume it holds for all~$r$, in which case $c_i$ is equal to the expected degree of node~$i$ and $m$ is the expected number of edges in the network.  At the same time, the expected number of edges falling between nodes in group~$r$ and group~$s$ is
\begin{equation}
    m_{rs} = \sum_{ij} {c_i c_j\over2m} \omega_{rs} \delta_{rg_i} \delta_{sg_j}
    = {\kappa_r\kappa_s\over2m} \omega_{rs}.
    \label{eq:mrs_microcanonical}
\end{equation}

Now let $p_r(c)$ be the probability density of values of~$c_i$ within group~$r$, let $\av{c}_r = \int_0^\infty c p_r(c) \>\dd c$ be its average, and let $q_r(c) = cp_r(c)/\av{c}_r$ be the corresponding excess distribution.  Then the fundamental generating functions for each group are given by the Laplace transforms
\begin{equation}
g^{(r)}_0(z) = \mathcal{L}\{p_r\}(1-z), \quad g^{(r)}_1(z) = \mathcal{L}\{q_r\}(1-z).
\label{eq:canonical_g}
\end{equation}
Now we solve the microcanonical model with these generating functions.  The number of edges between groups is drawn from a Poisson distribution with mean given by Eq.~\eqref{eq:mrs_microcanonical}, but the solution of the microcanonical model only requires the ratio~$m_{rs}/\kappa_r$, as in Eq.~\eqref{eq:u_dcsbm}, which becomes tightly concentrated about its mean as the network becomes large, so it suffices to solve the equations using this mean value and all the results of Section~\ref{sec:microcanonical} for the giant component, small components, and other features of the network carry over without modification.

As an example, consider a canonical block model with $K$ equally sized groups of $n/K$ nodes each and an exponential distribution of expected degrees in each group:
\begin{equation}
    p_r(c) = {1\over\mu_r} e^{-c/\mu_r},\qquad
    q_r(c) = {c\over\mu_r^2} e^{-c/\mu_r},
    \label{eq:canonical_pq}
\end{equation}
where $\mu_r$ is the average degree within group~$r$.  Then the generating functions of Eq.~\eqref{eq:canonical_g} are
\begin{equation}
    g^{(r)}_0(z) = {1\over1+\mu_r(1-z)}, \quad
    g^{(r)}_1(z) = {1\over[1+\mu_r(1-z)]^2},
    \label{eq:canonical_exp}
\end{equation}
and $\kappa_r = n\mu_r/K$, and Eq.~\eqref{eq:u_dcsbm} becomes
\begin{equation}
    u_r = {1\over Kc} \sum_s {\omega_{rs}\mu_s\over[1+\mu_s(1-u_s)]^2},
\end{equation}
where $c$ is the overall average degree of the network and we have used Eq.~\eqref{eq:mrs_microcanonical} for~$m_{rs}$.  This gives us $K$ simultaneous cubic equations, which we can solve iteratively.  In fact, however, we don't need to do that in this case because the form of the generating functions in~\eqref{eq:canonical_exp} is functionally the same as that for the microcanonical model with geometric degree distribution in Eq.~\eqref{eq:geometric_g}, and hence, for suitable choice of parameter values the behavior of the two models is the same: the phase diagram will look like Fig.~\ref{fig:phasediag} and the distribution of component sizes will look like Fig.~\ref{fig:small_components}.

\section{Percolation}
\label{sec:percolation}
The same techniques we use for determining component structure of block model networks can also be used, with some modification, to determine percolation properties of the same networks for either node or edge percolation~\cite{CNSW00,CEBH01}.  In this section we demonstrate these developments and again consider our three models in turn, starting with the ordinary (not degree-corrected) stochastic block model.

\subsection{Ordinary stochastic block model}
Consider a percolation process on the ordinary stochastic block model in which either nodes or edges are occupied uniformly at random with probability~$\phi$.  (The occupation probability is more commonly denoted~$p$, but we use~$\phi$ to avoid confusion with other quantities called~$p$ in this paper.)  Provided the network contains a giant component, a giant percolating cluster is guaranteed to form as $\phi$ increases, with a percolation transition at some value~$\phi_c$.  Let $u_r$ be the probability that a node in group~$r$ is \emph{not} connected to the percolating cluster via any of its incident edges.  Then for the case of node percolation the fraction of nodes in group~$r$ belonging to the giant cluster is given by
\begin{equation}
  S_r = \phi(1 - u_r),
  \label{eq:Sr_node1}
\end{equation}
while for edge percolation it is given by
\begin{equation}
  S_r = 1 - u_r.
  \label{eq:Sr_edge1}
\end{equation}
(The extra factor of~$\phi$ in the node percolation case accounts for the requirement that the node itself must be occupied if it is to belong to the percolating cluster.)

Now consider the probability that a node in group~$r$ is not connected to the percolating cluster via a specific other node that is in group~$s$.  There are several ways this could happen.  First, there could be no edge at all between the two nodes, which happens with probability $1-\omega_{rs}$.  Second, there could be an edge (probability~$\omega_{rs}$) but that edge could be unoccupied (for edge percolation) or the node at its end could be unoccupied (for node percolation), both of which happen with probability $1-\phi$.  Or third, the edge could exist and be occupied (probability~$\phi\omega_{rs}$), but the node at its end is not itself in the percolating cluster (probability~$u_s$).  Putting things together the total probability is
\begin{equation}
    1 - \omega_{rs} + (1-\phi)\omega_{rs} + \phi\omega_{rs} u_s
    = 1 - \phi\omega_{rs} (1-u_s).
\end{equation}
Taking a product over all nodes then gives us the complete probability~$u_r$ thus:
\begin{align}
      u_r &= \prod_s \bigl[1 - \phi\omega_{rs} (1-u_s)\bigr]^{n_s}
       = \prod_s \biggl[1 - \phi {c_{rs}\over n_s} (1-u_s)\biggr]^{n_s} \nonumber\\
       &\simeq e^{-\phi \sum_s c_{rs}(1 - u_s)},
       \label{eq:perc_SBM}
\end{align}
where $c_{rs}$ is defined as in Eq.~\eqref{eq:crs} and the final equality becomes exact in the limit of large network size.  Equation~\eqref{eq:perc_SBM} is similar to, and generalizes, Eq.~\eqref{eq:gc_SBM}, and the two become the same when $\phi=1$ so that all nodes or edges are occupied.  Now using Eqs.~\eqref{eq:Sr_node1} and~\eqref{eq:Sr_edge1} we have
\begin{equation}
    S_r = \phi\Bigl( 1 - e^{-\sum_s c_{rs} S_s} \Bigr)
    \label{eq:Sr_node2}
\end{equation}
for node percolation or
\begin{equation}
    S_r = 1 - e^{-\phi\sum_s c_{rs} S_s}
    \label{eq:Sr_edge2}
\end{equation}
for edge percolation, and the size of the complete percolating cluster is
\begin{equation}
    S = {1\over n} \sum_r n_r S_r
    \label{eq:Scomplete}
\end{equation}
in either case.

Equations~\eqref{eq:Sr_node2} and~\eqref{eq:Sr_edge2} do not have closed-form solutions for $S_r$ but can be solved by simple iteration starting from any reasonable initial values.  If there is no percolating cluster in the network then necessarily this iteration converges to $S_r=0$ for all~$r$.  If there is a percolating cluster it will converge to some nonzero solution.  Thus, as with Eq.~\eqref{eq:gc_SBM}, the question of whether the network has a percolating cluster is a question of whether the point $S_r=0$ is a stable fixed point of the iteration.  Expanding around this point in the regime where $S_r$ is small, we have
\begin{equation}
    S_r = \phi \sum_s c_{rs} S_s + \Ord(S^2),
\end{equation}
for either node or edge percolation and hence the iteration is unstable, and a percolating cluster exists, if and only if the leading eigenvalue of the Jacobian matrix~$\phi\mat{C}$ is greater than one, where $\mat{C}$ is the matrix of elements~$c_{rs}$ as previously.  Equivalently, there is a percolating cluster if $\phi$ is greater than $1/\lambda$, where $\lambda$ is the leading eigenvalue of~$\mat{C}$, and hence the percolation threshold falls at
\begin{equation}
    \phi_c = {1\over\lambda}.
\end{equation}

\begin{figure}
\begin{center}
\includegraphics[width=\linewidth]{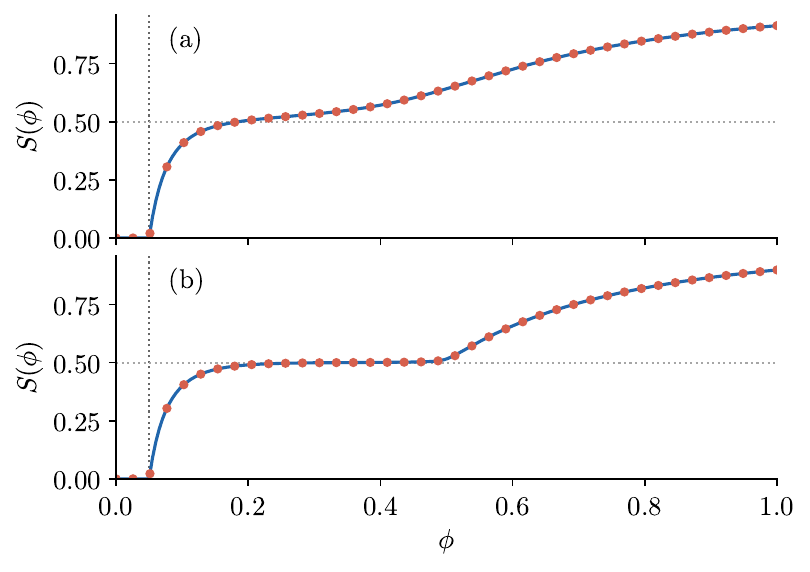} 
\end{center}
\caption{Size of the percolating cluster as a function of edge occupation probability for edge percolation on two different stochastic block models.  Both models have two equally-sized groups with $c_{11} = 2$ and $c_{22} = 20$ and the other parameters are (a)~$c_{12} = c_{21} = 0.1$ and (b)~$c_{12} = c_{21} = 0.001$.  The solid curves represent the exact solutions in the limit of infinite network size, from Eq.~\eqref{eq:Scomplete}, while the points are measurements of the largest component size averaged over 20 realizations of the percolation process on networks with 500\,000 nodes each.  The horizontal dotted line is at $S = \frac12$, corresponding to half of the nodes belonging to the giant cluster, and the vertical dotted line is at the critical occupation probability $\phi_c = 1/\lambda$.}
\label{fig:sbm_percolation}
\end{figure}

Figure~\ref{fig:sbm_percolation} shows two example applications to networks with two groups of equal size and different choices of parameters.  The plots show the size of the giant component for edge percolation for each case, calculated using Eq.~\eqref{eq:Sr_edge2}, along with direct simulation results on randomly generated networks from the same models, calculated using the fast algorithm of~\cite{NZ00,NZ01}.  As the figure shows, agreement between theory and simulations is good.

An interesting feature of this particular example is the ``double jump'' shape of the curve visible particularly in panel~(b) of the figure, though also to a lesser extent in panel~(a).  This kind of behavior arises in cases such as this one where the network is strongly assortative, meaning that the probability of between-group edges is particularly low.  In this situation we almost have two separate networks, each with its own percolation transition, visible as the two jumps in the curve.  For networks with more than two groups we can have multiple such jumps, one for each group.

\subsection{Microcanonical block model}
Turning next to the microcanonical version of the degree-corrected block model, we consider a similar edge or node percolation process with occupation probability~$\phi$ and let $u_r$ now represent the probability that a node in group~$r$ is not connected to the percolating cluster via a specific neighboring node.  There are two ways this can happen.  First, the edge in question or the node at its other end can be unoccupied (depending on whether we are looking at edge or node percolation) or, second, it can be occupied but the node at the other end is itself not connected to the percolating cluster.  The combined probability of these two outcomes is $1 - \phi + \phi u_s^k$, where $k$ is the excess degree of the neighboring node and $s$ is the group to which it belongs.  Averaging over the excess degree distribution then gives us an average probability of
\begin{equation}
    \sum_k q^{(s)}_k \bigl[ 1 - \phi + \phi u_s^k \bigr]
    = 1 - \phi + \phi g^{(s)}_1(u_s).
\end{equation}
The probability of the edge connecting to a node in group~$s$ in the first place is $m_{rs}/\kappa_r$ as previously, and hence, averaging over all groups, we have
\begin{equation}
u_r = {1\over\kappa_r} \sum_s m_{rs} \bigl[ 1 - \phi + \phi g^{(s)}_1(u_s) \bigr].
\label{eq:u_percolation_dcsbm}
\end{equation}
There is one equation~\eqref{eq:u_percolation_dcsbm} for each group and we can solve the complete set by simple iteration, then the probability~$S_r$ that a node in group~$r$ is in the percolating cluster is
\begin{equation}
    S_r = \phi\bigl[ 1 - g^{(r)}(u_r) \bigr]
\end{equation}
for node percolation or
\begin{equation}
    S_r = 1 - g^{(r)}(u_r)
\end{equation}
for edge percolation.  And the overall probability of being in the complete percolation cluster, which is also the expected size of the cluster, is given by Eq.~\eqref{eq:Scomplete} again.

If there is no percolating cluster then, as before, the iteration necessarily converges to $u_r=1$, while if there is a percolating cluster it will converge to some other solution, so the stability of the $u_r=1$ point tells us when such a cluster exists.  Setting $u_r = 1 - \epsilon_r$ once more and expanding in the small quantity~$\epsilon_r$, Eq.~\eqref{eq:u_percolation_dcsbm} becomes
\begin{equation}
    \epsilon_r = {\phi\over\kappa_r} \sum_s m_{rs} \biggl[ \biggl( {\dd g^{(s)}_1\over\dd z} \biggr)_{z=1} \epsilon_s + \Ord(\epsilon_s^2) \biggr].
\end{equation}
Neglecting the sub-leading terms, this implies that $\epsilon_r$ will grow under the iteration, and hence there is a percolating cluster, if and only if the leading eigenvalue of the Jacobian matrix~$\mat{J}$ is greater than~1, where the elements of the matrix in this case are
\begin{equation}
J_{rs} = \phi{m_{rs}\over\kappa_r} \tilde{c}_s,
\end{equation}
where $\tilde{c}_r$ is the average excess degree in group~$r$ as previously (see Eq.~\eqref{eq:tildec}).  Equivalently, we can define a matrix $\boldsymbol{\Lambda}$ with elements
\begin{equation}
\Lambda_{rs} = {m_{rs}\over\kappa_r} \tilde{c}_s
\end{equation}
and leading eigenvalue~$\lambda$, then the percolation threshold, for either node or edge percolation, falls at critical occupation probability
\begin{equation}
    \phi_c = {1\over\lambda}.
\end{equation}

We can extend these calculations to also solve for the size distribution of small clusters.  Let $h^{(r)}_1(z)$ be the generating function for the number of nodes reachable along an edge in group~$r$ as before.  If the edge is unoccupied (for edge percolation) or the node at its end is unoccupied, then this number is zero.  If the node or edge is occupied then as previously the number has generating function $z g^{(s)}_1(h^{(s)}_1(z))$, where $s$ is the group to which the node at the other end belongs.  Combining these observations and averaging over all groups~$s$, we then have
\begin{equation}
    h^{(r)}_1(z) = {1\over\kappa_r} \sum_s m_{rs} \bigl[ 1 - \phi + \phi z g^{(r)}_1\bigl( h^{(r)}_1(z) \bigr) \bigr].
    \label{eq:h1_percolation_dcsbm}
\end{equation}
By a similar argument the generating function for the size of the small cluster to which a node in group~$r$ belongs is
\begin{equation}
    h^{(r)}_0(z) = 1 - \phi + \phi z g^{(r)}_0\bigl( h^{(r)}_1(z) \bigr)
\end{equation}
for node percolation or
\begin{equation}
    h^{(r)}_0(z) = z g^{(r)}_0\bigl( h^{(r)}_1(z) \bigr)
    \label{eq:h0_percolation_dcsbm}
\end{equation}
for edge percolation.  From these one can derive the full size distribution either by finite iteration or by a Fourier transform, as in Section~\ref{sec:small}.

\begin{figure}[t]
    \begin{center}
    \includegraphics[width=\linewidth]{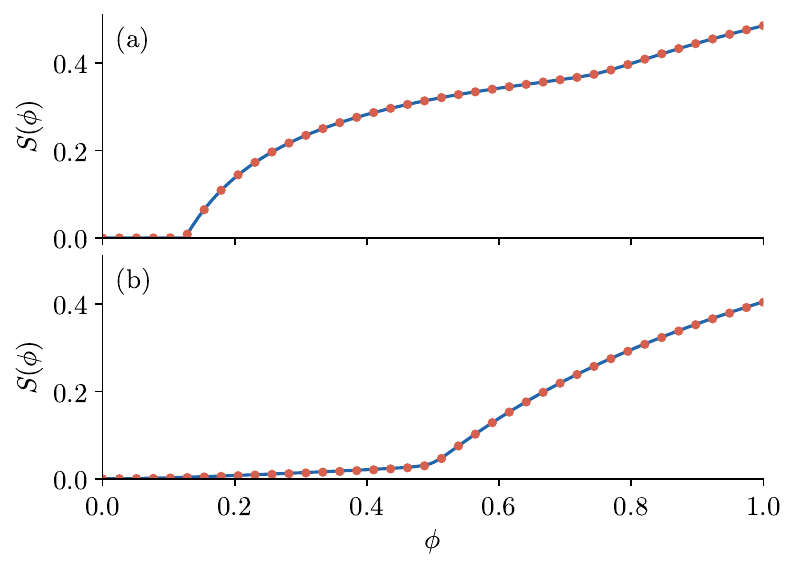}
    \end{center}
\caption{Size of the giant cluster as a function of edge occupation probability for edge percolation on two microcanonical block models. The solid lines represent the exact solution, while the points are measurements of the largest component size averaged over twenty realizations of the percolation process.  (a)~A network with two equally sized groups with geo\-metric degree distributions $p^{(r)}_k \sim a^k$, with parameter~$a$ equal to 0.5 in one group and 0.95 in the other.  (b)~A network with two groups of sizes $n_1 = 0.1n$ and $n_2 = 0.9n$, the first group with a power-law degree distribution with exponent $\alpha = 2.5$ and the second with a geometric degree distribution with $a = 0.5$.  Both networks have $m_{12}/\kappa_1 = 0.001$ and $n=500\,000$.}
  \label{fig:dcsbm_percolation}
\end{figure}

Figure~\ref{fig:dcsbm_percolation} shows two example applications for edge percolation on degree-corrected model networks with two groups.  The first example has two groups of equal size~$\frac12 n$ and node degrees distributed according to two different geometric distributions.  The second example consists of a small group of size $0.1n$ with a power-law degree distribution and a large group of size $0.9n$ with a geometric distribution.  The latter example reveals an interesting effect.  As shown elsewhere~\cite{CNSW00,CEBH01}, the percolation threshold is at $\phi=0$ for a power-law network, so the power-law group in our second network always has a percolating cluster and hence so does the complete network, although the size of the cluster as a fraction of the system size is relatively small because the group is small.  Only at a substantially larger value of $\phi$ around 0.5 does a percolating cluster appear in the other group, at which point the overall size of the percolating cluster in the whole network starts to grow rapidly.  This phenomenon implies that any group of nodes with a power-law degree distribution in a network, even a small group, is enough to ensure that the network will always have a percolating cluster, although that cluster may be small until percolation starts to also occur elsewhere in the network.

\subsection{Canonical block model}
Given our solution of the percolation problem for the microcanonical model, the equivalent calculation for the canonical model is now straightforward using the methods of Section~\ref{sec:canonical}.  As described in that section, the canonical model is equivalent to a microcanonical model with an appropriately modified degree distribution, which means that the fundamental generating functions $g^{(r)}_0(z)$ and $g^{(r)}_1(z)$ for the degree distribution and excess degree distribution in group~$r$ are replaced with Laplace transforms $\mathcal{L}\{p_r\}(1-z)$ and $\mathcal{L}\{q_r\}(1-z)$ as in Eqs.~\eqref{eq:replace0} and~\eqref{eq:replace1}.  Then, for example, the fraction of nodes in group~$r$ that belong to the percolating cluster can be derived from the solution of the equations
\begin{equation}
u_r = {1\over\kappa_r} \sum_s m_{rs} \bigl[ 1 - \phi + \phi \mathcal{L}\{q_s\}(1-u_s) \bigr],
\end{equation}
and
\begin{equation}
    S_r = \phi\bigl[ 1 - \mathcal{L}\{p_r\}(1-u_r) \bigr]
\end{equation}
for node percolation or
\begin{equation}
    S_r = 1 - \mathcal{L}\{p_r\}(1-u_r)
\end{equation}
for edge percolation.  And the size of the complete percolating cluster is given by Eq.~\eqref{eq:Scomplete} as before.  Equations for the distribution of small component sizes can also be derived by making the same replacements of the generating functions in Eqs.~\eqref{eq:h1_percolation_dcsbm} to~\eqref{eq:h0_percolation_dcsbm}.

\subsection{Nonuniform occupation probability}
So far we have looked at traditional percolation processes in which nodes or edges are occupied uniformly and independently, but there are other variants of percolation that could be studied on block model networks.  As an example, consider percolation in which nodes or edges are again occupied independently but now with nonuniform probabilities.  The most common case, and the only one widely studied in the past, is node percolation in which the occupation probability~$\phi_k$ is some function of node degree~$k$.  For example if $\phi_k=1$ for $k\le k_\text{max}$ and 0 otherwise, then we effectively remove from the network all nodes with degree greater than~$k_\text{max}$, which can be a model for targeted attack on a infrastructure network or selective vaccination of highly-exposed individuals in a contact network.

Degree-dependent percolation like this can be treated with similar techniques to the uniform case.  For instance, consider the microcanonical degree-corrected model and let us calculate the size of the giant percolating cluster.  We define $u_r$ to be the probability that an edge from a node in group~$r$ does not connect to the percolating cluster, so that if the node has degree~$k$ then $u_r^k$ is the probability that it is not connected to the percolating cluster via any of its edges and $1-u_r^k$ is the probability that it is connected.  In order to be in the percolating cluster the node must also itself be occupied, which happens with probability~$\phi_k$, so the overall probability of being in the percolating cluster is $\phi_k (1-u_r^k)$.  Averaging over the degree distribution~$p^{(r)}_k$, we then have an average probability~$S_r$ of
\begin{equation}
    S_r = \sum_k p^{(r)}_k \phi_k (1-u_r^k) = f^{(r)}_0(1) - f^{(r)}_0(u_r),
\end{equation}
where we have defined a new generating function
\begin{equation}
    f^{(r)}_0(z) = \sum_k p^{(r)}_k \phi_k z^k.
\end{equation}
Then the size of the complete percolating cluster is given once again by Eq.~\eqref{eq:Scomplete}.

We can calculate $u_r$ by a similar process.  Suppose an edge from group~$r$ connects to a node in group~$s$ with excess degree~$k$.  Then the probability of the latter node being occupied is $\phi_{k+1}$ (because the excess degree is one less than the total degree) and the probability that the edge does not connected to the percolating cluster is $1-\phi_{k+1}+\phi_{k+1} u_s^k$.  Averaging over both the excess degree distribution and all groups~$s$ we then get
\begin{align}
    u_r &= {1\over\kappa_r} \sum_s m_{rs} \sum_k q^{(s)} \bigl[ 1 - \phi_{k+1} + \phi_{k+1} u_s^k ] \nonumber\\
    &= {1\over\kappa_r} \sum_s m_{rs} \bigl[ 1 - f^{(s)}_1(1)] - f^{(s)}(u_s) \bigr],
\end{align}
where
\begin{equation}
    f^{(r)}_1(z) = \sum_k q^{(r)}_k \phi_{k+1} z^k.
\end{equation}

\begin{figure}[t]
  \begin{center}
  \includegraphics[width=\linewidth]{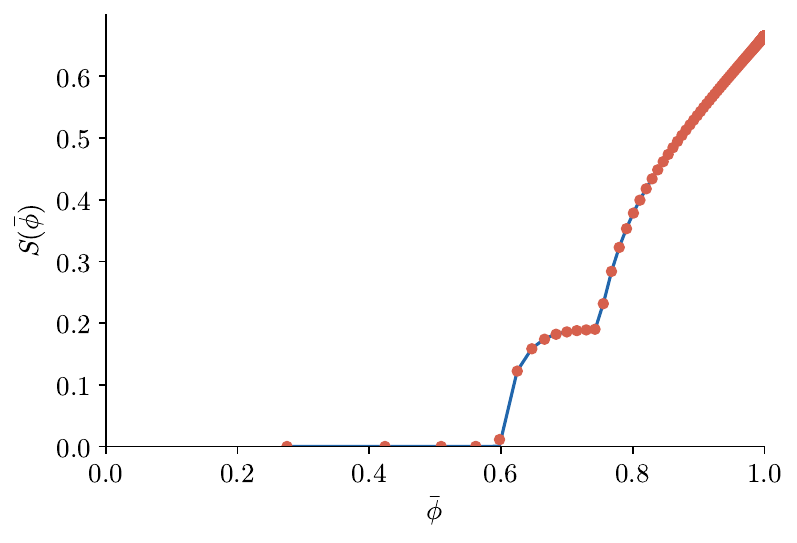}
  \end{center}
  \caption{Size of the percolating cluster as a function of the fraction of occupied nodes as the highest degree nodes are removed, for a microcanonical block model with 200\,000 nodes and two equally-sized groups with geometric degree distributions with parameters $a = 0.5$, $0.95$ and mixing parameter $m_{12}/\kappa_1 = 0.999$.  The solid lines represent the exact solution while the points are measurements of the largest component size on computer generated networks.}
  \label{fig:targeted_percolation}
\end{figure}

Consider for example the case discussed above where all nodes with degree greater than~$k_\text{max}$ are removed from the network by setting $\phi_k$ to zero.  The size of the percolating cluster, calculated from Eq.~\eqref{eq:Scomplete}, is shown in Fig.~\ref{fig:targeted_percolation} for a two-group network, as a function of the average fraction of occupied nodes
\begin{equation}
    \bar\phi = {1\over n} \sum_r n_r \sum_k p^{(r)}_k \phi_k
    = \sum_r {n_r\over n} f^{(r)}_0(1).
\end{equation}
The plot shows a number of interesting features.  Preferentially removing the high-degree nodes has the effect of increasing the percolation threshold because the network is sparser for given~$\bar\phi$ than an equivalent network with nodes removed uniformly.  This means that the network fails to percolate relatively quickly as nodes are removed (reading the plot from right to left), and hence that the network is fragile to targeted removal of its nodes, as noted previously for various other networks~\cite{AJB00,CNSW00,CEBH01}.  We also observe that the percolating cluster forms first in the group with lower average degree (which is the opposite of what happens for uniform percolation) because more nodes are removed from the group with higher average degree.

\section{Conclusions}
\label{sec:conclusions}
In this paper we have studied the component structure and percolation properties of community-structured networks drawn from the stochastic block model and its degree-corrected variants.  We have derived conditions for the existence of a giant component in the ordinary stochastic block model and the size of the giant component when there is one, and we have extended these results to the microcanonical version of the degree-corrected block model as well.  We have also derived a system of equations for the size distribution of small components in the latter model and presented two methods for solving them numerically.  We have then extended these results to the more widely studied canonical degree-corrected model, showing that the corresponding quantities in this model can be calculated from the solution for the microcanonical model by a straightforward replacement of the generating functions for degree distributions with Laplace transforms.  As a corollary, this approach also provides a quick method to derive the component structure of the Chung--Lu model, a canonical variant of the (microcanonical) configuration model.

We have also studied node and edge percolation on the same selection of models and derived expressions for the position of the percolation threshold, size of the percolating cluster when there is one, and complete distribution of small cluster sizes.  We have also looked briefly at the case of nonuniform percolation in which nodes are occupied with probability that depends on their degree.

The work presented in this paper could be extended in a number of directions.  Our methods could be used to study specific examples such as models with particular degree distributions---power-law distributions say, or other long-tailed distributions.  Our results for percolation could be used as the basis for understanding epidemiology on community structured networks by employing the known correspondence between percolation and the SIR model~\cite{MN00a,PV01a,Sander02,Newman02c}.  More broadly, one could also study other percolation-like processes on block models, including complex contagion processes or dynamical processes, or extend the methods of this paper to more elaborate network models such as overlapping or mixed-membership models.

\end{document}